\documentclass[
    ,final            
    ,numberedheadings 
  ]
  {aipproc}

\layoutstyle{6x9}

\usepackage{amssymb}
\usepackage{graphicx}        
\usepackage{bbm}
\usepackage[small,loose]{subfigure}
\usepackage[centertags]{amsmath}
\usepackage{epsfig}
\usepackage{epic}
\usepackage{eepic}
\usepackage{color}

\newcommand{\BmL}{\ensuremath{B\!-\!L} }
\newcommand{\PQ}{\ensuremath{\mathrm{U(1)}_\mathrm{PQ}}}
\newcommand{\E}[1]{\ensuremath{\mathrm{E}_{#1}}} 

\newcommand{\SO}[1]{\ensuremath{\mathrm{SO}(#1)}}
\newcommand{\SU}[1]{\ensuremath{\mathrm{SU}(#1)}}
\newcommand{\U}[1]{\ensuremath{\mathrm{U}(#1)}}
\newcommand{\x}{\ensuremath{\times}}
\newcommand{\Z}[1]{\ensuremath{\mathbbm{Z}_{#1}}} 
\newcommand{\I}{\mathrm{i}}

\begin{document}

\begin{flushright}
\normalsize{DESY-09-147}\\
\normalsize{LMU-ASC 40/09}\\
\end{flushright}

\title{Local Grand Unification and String Theory}


\author{Hans Peter Nilles}{
  address={Bethe Center for Theoretical Physics and Physikalisches Institut der Universit\"at Bonn,\\ Nussallee 12, 53115 Bonn, Germany}
}
\author{Sa\'ul Ramos-S\'anchez}{
  address={Deutsches Elektronen-Synchrotron DESY, Hamburg, Germany}
}
\author{Patrick K.~S. Vaudrevange}{
  address={Arnold Sommerfeld Center for Theoretical Physics, Ludwig-Maximilians-Universit\"at M\"unchen, 80333 M\"unchen, Germany}
}

\begin{abstract}
  The low energy effective action of string theory depends strongly on the process of compactification
  and the localization of fields in extra dimensions. Explicit string constructions towards the minimal
  supersymmetric standard model (MSSM) reveal interesting results leading to the concept of local grand
  unification. Properties of the MSSM indicate that we might live at a special location close to an
  orbifold fixed point rather than a generic point in Calabi-Yau moduli space. We observe an enhancement
  of (discrete) symmetries that have various implications for the properties of the MSSM such as proton
  stability as well as solutions to the flavor problem, the $\mu$-problem and the strong CP-problem.
\end{abstract}

\maketitle


\section{Introduction}

There are good reasons to believe that all fundamental forces allow for a unified description.  The running
of the gauge couplings of strong and electroweak interactions and the symmetries of the particle content 
of the standard model (SM) are the most compelling ones and suggest a unified picture of these 
interactions through grand unified theories (GUTs)~\cite{Georgi:1974sy,Pati:1974yy}. The fundamental 
feature of these theories is that, at some high scale $M_{GUT}$, all gauge interactions of the SM 
are described by a single and bigger gauge group, such as \SU5, \SO{10} or \E6 with a single 
(unified) gauge coupling. However, precision measurements at the weak scale indicate that within 
the SM the gauge couplings do not meet accurately. The situation is improved if supersymmetry (susy) 
is included. In the minimal supersymmetric extension of the SM (the MSSM), all gauge couplings meet 
(with acceptable accuracy) at $M_{GUT}\sim3\x10^{16}$ GeV. Beside gauge coupling unification, 
GUTs offer natural explanations for neutrino masses, Yukawa unification and the structure of the matter 
generations.

A particularly interesting GUT is \SO{10}~\cite{Georgi:1975qb,Fritzsch:1974nn}. In this theory, one
generation of quarks and leptons is elegantly accommodated within a single spinor representation of \SO{10},
according to the decomposition
\begin{equation}
\label{eqn:16plet}
\mbox{
{\footnotesize
$\displaystyle
  \begin{array}{rccccccccccc}
    \mathbf{16}~=\!&\!(\mathbf{3}, \mathbf{2})_{1/6} &\!\oplus\!& (\mathbf{\overline{3}}, \mathbf{1})_{-2/3} &\oplus&
    (\mathbf{\overline{3}}, \mathbf{1})_{1/3} &\oplus&
    (\mathbf{1}, \mathbf{2})_{-1/2} &\oplus& (\mathbf{1}, \mathbf{1})_{1} &\oplus& (\mathbf{1}, \mathbf{1})_{0} \;,\\
    &q &&\overline{u} &&\overline{d} && \ell && \overline{e} && \overline{\nu}
  \end{array}
$
}}
\end{equation}
where quantum numbers with respect to $G_{SM}=\SU3_c\x\SU2_L\x\U1_Y$ are displayed.
Remarkably, \SO{10} GUTs predict the existence of right--handed neutrinos $\overline{\nu}$ and include an 
additional $\U1$ (named $\U1_{\BmL}$) that forbids dangerous dimension 4 proton decay operators.

Beside their attractive properties, GUTs introduce some problems of their own. The most puzzling feature is
that, while matter generations are described by complete GUT representations (eq.~\eqref{eqn:16plet}), 
SM Higgs and gauge bosons appear only as {\it split} (incomplete) GUT multiplets. Hence, additional 
fields are needed in order to obtain full GUT representations. However, these fields are problematic as they 
generically mediate fast proton decay. In the case of the SM Higgs, this is known as the {\it doublet-triplet 
splitting problem} and is present in all interesting GUTs. Apart from that, the breaking of GUT groups down 
to $G_{SM}$ is rather involved and requires additional Higgs fields in large representations (e.g. 
$\mathbf{\overline{126}}$ of \SO{10}). 

Some of these issues can be solved in higher-dimensional field theories compactified on orbifolds. 
Starting with a GUT group in 5 or 6D, the GUT symmetry breaking to $G_{SM}$ is not induced by a Higgs 
mechanism, but by choosing appropriate boundary conditions for the gauge bosons in the extra dimension(s). 
Furthermore, by placing the three SM generations and the Higgs fields on different brane-like objects in 
the extra dimension(s), i.e. on fixed points of the orbifold, one can find a geometrical explanation for 
the difference between complete GUT matter representations and split Higgs 
multiplets~\cite{Kawamura:2000ir,Asaka:2001eh,Asaka:2003iy}.

These features studied in field theory orbifolds are also proper to compactifications of string theory. In 
particular, the chain of GUTs
\begin{equation}
  \label{eq:GUTschain}
   G_{SM}\;\subset\;\SU5\;\subset\;\SO{10}\;\subset\;\E6\;\subset\;\E7\;\subset\;\E8
\end{equation}
suggests that \E8\x\E8 heterotic orbifolds~\cite{Dixon:1985jw,Dixon:1986jc} are natural candidates to
provide an ultraviolet complete theory that solves the inherent problems of GUTs. 

\begin{figure}[t]
  \noindent
  \begin{minipage}{0.45\linewidth}
    {\footnotesize a)}\\[3mm]
    \includegraphics[width=0.8\textwidth]{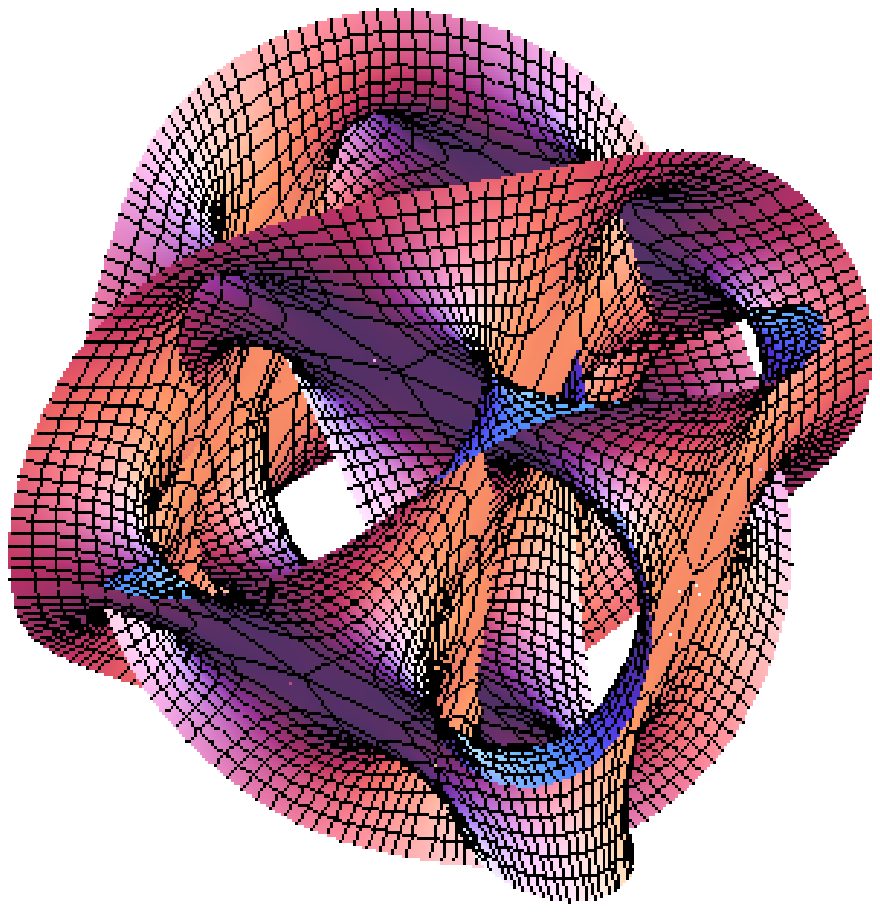}
  \end{minipage}
  \hspace{2mm}
  \begin{minipage}{0.4\linewidth}
    {\footnotesize b)}\\
    \includegraphics{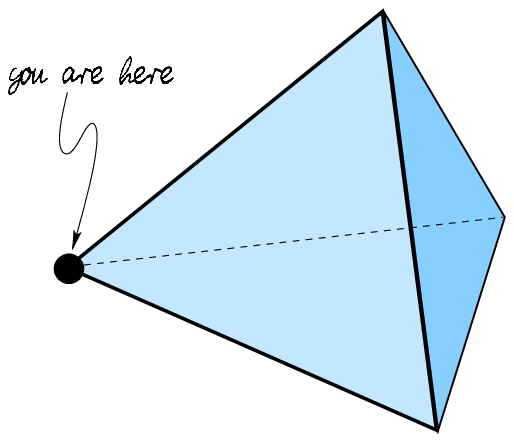}
  \end{minipage}
  \vspace{-0.3cm}
  \caption{a) Calabi-Yau spaces versus b) orbifolds. Orbifolds are singular limits of Calabi-Yau spaces
   with a much simpler structure. Whereas studying the phenomenological properties of Calabi-Yau compactifications 
   is very hard, many appealing features arise naturally in orbifolds suggesting that we might actually 
   live close to an orbifold fixed point rather than a generic point in
   Calabi-Yau moduli space~\cite{Nibbelink:2009sp}.}
\label{fig:orbiVsCY}
\end{figure}

Heterotic orbifolds can be seen geometrically as singular limits of 6D smooth Calabi-Yau spaces (see 
fig.~\ref{fig:orbiVsCY}) and lead to models that resemble the 
MSSM~\cite{Ibanez:1987sn,Kobayashi:2004ya,Buchmuller:2005jr,Kim:2007mt} (for details on their 
construction, see e.g.~\cite{Bailin:1999nk,Vaudrevange:2008sm,RamosSanchez:2008tn,Faraggi:1997dc}). In the transition 
to the orbifold, the complicated Calabi-Yau reduces to a simple space that is flat everywhere except 
for some isolated singularities, the so-called fixed points. This results in symmetry enhancements of 
two kinds: first, discrete symmetries with geometrical origin arise, and second, the gauge symmetry is 
enhanced. Both features are desirable from the SM perspective: discrete symmetries can provide answers to 
prevailing puzzles of low energy phenomenology, such as the strong CP-problem as discussed in 
section~\ref{sec:accions}, and the 4D SM gauge symmetry can be enhanced locally at the various 
singularities to various GUTs. As described in the next section, the last feature gives rise 
to the concept of {\it local GUTs}~\cite{Kobayashi:2004ya,Buchmuller:2005jr,Forste:2004ie}, which has 
proven to be useful in heterotic orbifolds and recently also in F-theory~\cite{Beasley:2008kw}.

\section{Local Grand Unification}
\label{sec:localGUTs}

The concept of local grand unification in heterotic orbifolds is based on the observation that the \E8\x\E8 
gauge symmetry of the 10D bulk is broken only at the orbifold singularities to various subgroups, giving rise to 
so-called local GUTs. In this section, we discuss this concept in more detail and show that local GUTs seem 
to be a necessary link between the heterotic string and the MSSM. 

\subsection{Local GUTs in Heterotic Orbifolds}

Compactification of the heterotic string on orbifolds remains a simple and elegant method to achieve 4D
theories that include chiral fermions and phenomenologically viable gauge groups. Orbifolds emerge from
dividing a manifold by one of its discrete symmetries. We focus here on 6D toroidal abelian orbifolds, i.e.
on orbifolds resulting from moding a discrete \Z{N} symmetry out of a 6D torus. The whole curvature of
the emerging space is concentrated at the points that are left fixed under the action of the generator
$\vartheta$ of the \Z{N} symmetry (from the 10D point of view, these points are brane-like objects of 4 
or 6 dimensions). Internal consistency of the theory demands $\vartheta$ to be
associated to an operation in the \E8\x\E8 gauge degrees of freedom which, in the bosonic formulation used
here, is encoded in a 16D shift vector $V$. In presence of Wilson
lines~\cite{Ibanez:1986tp}, the gauge embedding varies at different localizations in the compact space,
i.e. it is different from brane to brane. Thus, it is necessary to introduce the local shift
$V_{\rm local}$ that parametrizes the orbifold action on the branes.

In heterotic orbifolds, massless states originate from closed strings of two types. The untwisted or bulk
states are free to move in the whole space and stem directly from the strings associated to the 10D
supergravity and \E8\x\E8 vector multiplets. The 4D gauge bosons belong to this category. In addition, 
there are other states of pure stringy origin, the so-called twisted states.
The corresponding strings close only thanks to the \Z{N} symmetry, what
constrains them to be attached to the associated fixed points.  These states are matter fields 
in the 4D effective theory.

\begin{figure}[t]
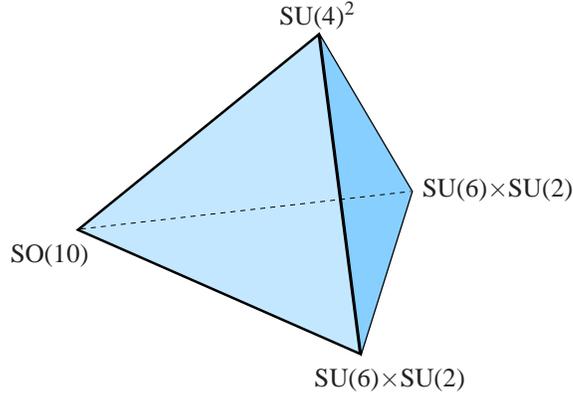

  \centerline{\input tetrahedral3.pstex_t}\vspace{-0.3cm}
  \caption{Gauge group topography of a \Z2 orbifold. In heterotic orbifold compactifications, the bulk
    gauge group is $\E8\x\E8$ and is broken at the singularities (corners) to local GUTs.  The 4D gauge
    group, i.e. the common intersection of the local ones, is in this example $\SU3_c\times\SU2_L\times\U1_Y$.}
\label{fig:tetra}
\end{figure}

In a second step, bulk and twisted states are affected by the action of the orbifold. They
acquire phases that depend on $\vartheta$ and its (local) gauge embedding $V_{\rm local}$. From the perspective 
of a fixed point, only those states that are invariant under the local orbifold action remain massless. 
Omitting the details for the sake of brevity, this implies that the only gauge
bosons surviving the local orbifold projection are those that satisfy
\begin{equation}
  \label{eq:gaugetrafo}
  e^{2\pi\I \ p \cdot V_{\rm local}}~=~1 \quad\Leftrightarrow\quad 
  p \cdot V_{\rm local}~=~0\mod 1\;,
\end{equation} 
where $p$ denotes the roots of the unbroken local gauge group.
It follows then that, at each fixed point of the orbifold there is a local gauge group 
$G_{\rm local}\subset\E8\x\E8$ that is larger than the one in 4D (see fig.~\ref{fig:tetra}). 
The effective low energy gauge group is the intersection of some or all $G_{\rm local}$, 
depending on the size of the internal space. For example, if all internal dimensions are 
of the order of the inverse string scale, the 4D gauge group is (in the absence of freely 
acting Wilson lines) the common subgroup to all local groups,
\begin{equation}
\label{eq:4Dgg}
G_{4D}~=~\bigcap_f ~G_{{\rm local},f}\;,
\end{equation}
where $f$ runs through all fixed points. On the other hand, if two dimensions are larger 
(of the order of the inverse GUT scale), the effective gauge group of the resulting 6D effective 
theory is given by the intersection of the local groups associated to the fixed points within the
four smaller dimensions. In some cases, this scenario can be interpreted as a 6D GUT which may 
break down to the SM in 4D. Another relevant observation is that twisted states located at 
the fixed points transform as complete representations of $G_{{\rm local},f}$, e.g. if the 
up-quark sits at some fixed point with local \SO{10} gauge group, it has to be accompanied by 
the other quarks and leptons to form a full $\mathbf{16}$--plet.

This favors \SO{10} local GUTs in models with $G_{4D}=G_{SM}$. If a $\mathbf{16}$--plet is 
found on a special GUT brane with local \SO{10} where the orbifold projection acts trivially, 
it appears in 4D as a complete generation of quarks and leptons. The Higgs fields, on the 
other hand, must not live on the same brane. They could be bulk states, transforming as 4D 
$\SU2_L$ doublets but not as full $\mathbf{10}$--plets of the local GUT.

\subsection{Fertile Patch in the Mini-Landscape}
\label{sec:MiniLandscape}

As soon as some 4D models with realistic properties were found from \Z6--II heterotic
orbifolds~\cite{Kobayashi:2004ya,Buchmuller:2005jr}, it was realized that they were endowed with local
GUTs. This seems to hint towards a selection principle: models with realistic properties are intimately
attached to local GUTs. If this principle is indeed true, finding the MSSM vacuum from string theory
reduces to spotting first the regions of the huge string
landscape~\cite{Susskind:2003kw,Schellekens:2006xz,Lust:2007kw} endowed with local GUTs.

How accurate the previous selection principle is can be judged only by taking it seriously as a guiding
strategy in the search for realistic vacua. This is the approach followed in the Mini-Landscape 
(ML)~\cite{Lebedev:2006kn,Lebedev:2007hv,Lebedev:2008un}, based on \Z6--II orbifolds (for details on \Z6--II, 
see e.g.~\cite{Kobayashi:2004ya,Buchmuller:2006ik}). The strategy is simple: one must focus on models endowed 
with local GUTs at special points where the orbifold projection is trivial and impose a series of 
phenomenological constraints. In the ML, we have demanded 
a) $G_{4D}=G_{SM}$ with local $\SO{10}$ or \E6 GUTs; 
b) gauge coupling unification ($\sin^2\theta_w = 3/8$) with 
nonanomalous hypercharge from $\SU5$;
c) three SM matter generations; and 
d) vectorlike exotics $x_i$ whose masses arise from couplings to SM singlets that acquire VEVs in susy vacua.
The ML search results in $\mathcal{O}(100)$ promising MSSM models, supporting the strength of the local GUT strategy.

Naturally, one might argue that there are perhaps other random regions of the \Z6--II string landscape far from
the local GUT patch that yield comparable vacua. In order to discard this, we have performed independently a 
random search~\cite{Lebedev:2008un}. The results are listed in table~\ref{tab:LocalGUTStructurePromisingModels}. 
First, out of a total of $\sim10^7$ models we have identified more than 200 hundred MSSM candidates. Second, we 
notice that most of the MSSM candidates have two Wilson lines, resulting in a $D_4$ family symmetry where two 
generations form a doublet and the third a $D_4$ singlet. Third, most of the promising models contain \E6 or 
\SO{10} or \SU5 local GUTs.

\begin{table}[t]
\centerline{
\begin{tabular}{ccrr}
\hline
    \tablehead{1}{c}{t}{Local GUT}
  & \tablehead{1}{c}{t}{``Family''}
  & \tablehead{1}{c}{t}{2 Wilson\\lines}
  & \tablehead{1}{c}{t}{3 Wilson\\lines}   \\
\hline
\E6          & $\mathbf{27}$ & $14$  & $53$ \\
\SO{10}      & $\mathbf{16}$ & $87$  &  $7$ \\
\SU6         & $\mathbf{15}+\mathbf{\overline6}$ &  $2$  &  $4$ \\
\SU5         & $\mathbf{10}$ & $51$  & $10$ \\
non GUTs &                  & $39$  &  $0$ \\
\hline
total &                 & $193$ & $74$ \\
\hline
\end{tabular}
\caption{Local GUT structure  of the  MSSM candidates. The \SU5 local GUT
does not  produce a complete family, so additional ``non-GUT'' states are
required.}
\label{tab:LocalGUTStructurePromisingModels}
}
\end{table}

These results indicate that a search strategy based on local GUTs leads to the vast majority of the
phenomenologically viable models. That is, as opposed to a random search, local GUTs are ideal
benchmarks to pinpoint the regions of the string landscape where the vacuum that describes our universe
may be found.

\paragraph{Supersymmetry Breaking, R--Parity and Seesaw Neutrinos}

The ML models have many further appealing features. First, it is found that susy is preserved at high
energies as long as there is a subset of SM singlets $s_i$ that acquire VEVs close to the string scale. 
This breaks many of the additional gauge factors in the 4D gauge group, but can leave
certain nonabelian hidden gauge groups unbroken. 
In the MSSM candidates, it turns out that these hidden sectors are appropriate for gaugino condensation 
and lead to susy breaking with $m_{3/2}\sim$ TeV~\cite{Lebedev:2006tr}.

In order to prevent the appearance of dangerous, dimension-four lepton ($L$) and baryon ($B$) number violating
operators that would allow for too rapid proton decay (e.g.  $\bar{u}\,\bar{d}\,\bar{d}$ and
$q\,\ell\,\bar{d}$), the MSSM includes R-parity (or matter parity)~\cite{Dimopoulos:1981dw}.  In some
ML models, matter parity could be identified from the spontaneous breaking of 
$\U1_{\BmL}\not\subset\SO{10}$ by the VEVs of some $s_i$ with even charges~\cite{Lebedev:2007hv},
\begin{equation}
  \label{eq:BmLtoZ2}
  \U1_{\BmL}\;\stackrel{s_i}{\longrightarrow}\;\Z2^{\rm matter}=(-1)^{3(\BmL)}\;.
\end{equation}
Notice that the existence of such singlets is associated with the fact that $\U1_{\BmL}$ turns out 
to be a mixture of the standard $B-L$ from \SO{10} and other \U1 factors. This symmetry allows the 
distinction between Higgs $h_d$ (even parity) and the leptons $\ell$ (odd), and between standard model 
singlets $s_i$ (even) and right--handed neutrinos $\overline{\nu}_j$ (odd). 
The promising ML models exhibit $\mathcal{O}(100)$ $\overline{\nu}_j$'s and the couplings
\begin{equation}
  \label{eq:neutrinocouplings}
  W~\supset~Y_\nu^{ij} \,h_u\,\ell_i\, \overline{\nu}_j + \frac{1}{2} M_{jk} \overline{\nu}_j\, \overline{\nu}_k \;,
\end{equation}
where $Y_\nu^{ij}$ and $M_{jk}$ are polynomials of the VEVs of $s_i$. This implies that the seesaw
mechanism is quite generic in orbifold models with realistic properties 
and the seesaw scale is slightly reduced due to the large number of right--handed neutrinos~\cite{Buchmuller:2007zd}.

Another generic feature of the ML MSSM candidates is the exceptional position of the top quark: 
only for the top quark there is a trilinear Yukawa coupling. In addition, this coupling originates from 
the \E8\x\E8 gauge coupling in 10D and hence the coupling strength $y_t$ is given by the gauge coupling 
$g$ at the unification scale. This equality receives corrections due to localization effects of bulk fields in the 
presence of localized Fayet-Iliopoulos terms that favor a slight suppression of $y_t$ against $g$~\cite{Hosteins:2009xk}.

From these properties, it seems natural to conclude that \Z6--II heterotic orbifolds furnished with local
GUTs represent a fertile framework for producing models connecting string theory with the MSSM.

\section{Accidental symmetries}

Similar to the appearance of accidental $B$ and $L$ number conservation in the standard model, global 
accidental symmetries can arise in string theory models as follows. 
10D string theory offers a wide range of exact symmetries: apart from gauge symmetries, there are (target-space) 
modular and continuous space-time symmetries. Due to the compactification to 4D, these symmetries are generically 
broken to a multitude of subgroups of various types. For example, for heterotic orbifolds the breaking of the 6D 
part of the 10D space-time symmetry results in some discrete (R-)symmetries~\cite{Kobayashi:2006wq,Araki:2008ek}. 
The unbroken symmetries restrict the form of the superpotential of the effective 4D theory and can be of great 
phenomenological relevance, e.g. as family symmetries~\cite{Ko:2007dz} or matter parity~\cite{Petersen:2009ip}. 
They could also provide an explanation of the stability of the proton~\cite{Dreiner:2005rd,Forste:2009xx}.

Additionally, focusing our attention on superpotentials of a limited degree in the fields, global accidental 
symmetries can arise~\cite{Weinberg:1972fn} that are explicitly broken to the exact discrete symmetries at higher 
orders in the superpotential. Nevertheless, some of these accidental symmetries, can remain unbroken up to 
very high orders and might be important to explain some of the issues of low energy physics such as the origin of 
large hierarchies in connection to the susy $\mu$--term problem~\cite{Kappl:2008ie}, moduli stabilization
or the strong CP-problem, as will be discussed in the following. 

\subsection{Accions}
\label{sec:accions}

The most elegant and appealing solution to the strong CP-problem is based on the conjecture of an axion
field~\cite{Weinberg:1977ma,Wilczek:1977pj}. 
It requires the existence of an anomalous global
Peccei-Quinn symmetry~\cite{Peccei:1977hh} \PQ\ and its spontaneous breakdown at a scale $F_a$ (where
$F_a$ denotes the axion decay constant). Constraints (mostly) from astrophysics and cosmology require
$F_a$ to be in the axion window
\begin{equation}
\label{eq:AxionBound}
10^9 \,{\rm GeV} \leq F_a \leq 10^{12} \,{\rm GeV}
\end{equation}
for the so-called ``invisible'' axion~\cite{Kim:1979if,Dine:1981rt}. The axion field adjusts its VEV
to cancel the $\theta$-parameter of quantum chromodynamics (QCD) to avoid
CP-violation due to strong interactions.

In the heterotic string, there are various sources for axions. The spontaneous breakdown of the anomalous
\U1 together with the Green-Schwarz mechanism produce the so-called model-independent axion, whose decay
constant is fixed by the Planck scale (see e.g.~\cite{Svrcek:2006yi}), too high to solve the strong
CP-problem. Model-dependent axions arise from the internal components of the $B-$field. Unfortunately,
their decay constants are in general not much lower than in the previous case~\cite{Svrcek:2006yi}.
Admissible axions, on the other hand, can arise from the breaking of (multiple) anomalous accidental
global \U1 symmetries realized as low energy remnants of (stringy) discrete
symmetries~\cite{Choi:2006qj}. These axion-like particles are referred to as {\it
  accions}~\cite{Choi:2009jt}.

The accion decay constant depends on the VEVs of the fields responsible for the breakdown of such global
\U1s. In general, if there are $N$ singlets acquiring VEVs and $M$ global accidental \U1s broken
spontaneously at hierarchically different scales, then  $F_a$ is of the scale of the $M$th largest VEV. 

It follows that in specific vacua in which some accidental \U1s are broken at an intermediate scale 
(e.g. $10^{12}$ GeV) a QCD accion decay constant satisfying current constraints can be achieved.
This situation is indeed realized in specific vacuum configurations of MSSM orbifold 
candidates~\cite{Choi:2009jt}. It has been additionally verified that the contributions to the accion
mass due to the explicit breaking of the accidental \U1s can lie in the admissible experimental window.

\section{Conclusions}

We have reviewed the concept of local GUTs and shown how models with local GUTs can alleviate some of
the disadvantages of traditional GUTs, such as the doublet--triplet splitting problem.
Moreover, this concept represents arguably one of the most
successful benchmarks to pinpoint the region of the immense string landscape where the vacuum that
describes our universe may be found. In particular, in the context of \Z6--II orbifolds, it leads to a
large set of MSSM candidates with the exact spectrum of the MSSM, coupling unification, seesaw neutrino 
masses and low energy susy breaking.

We have also studied some of the consequences of the various (global approximate) symmetries that appear
naturally in string compactifications as remnants of stringy discrete symmetries. These symmetries could
provide an explanation of the origin of the huge hierarchy between the Planck and electroweak scales.
Such hierarchies are also important for moduli stabilization. In addition, accidental symmetries lead to
natural candidates for QCD axions. 

As a general result, explicit string constructions towards the MSSM (and SM) seem to require local grand
unification and a very special location of fields in extra dimensions. We do not seem to live at a
generic point in Calabi-Yau moduli space but rather at a point of concentrated curvature that can be
described very well by an orbifold fixed point. Location of the fields close to that fixed point leads to
additional (discrete or accidental) symmetries which are relevant for specific properties of the MSSM.

\begin{theacknowledgments}
P.V. would like to thank LMU Excellent for support.
This research was supported by the DFG cluster of excellence 
Origin and Structure of the Universe, the
European Union 6th framework program  MRTN-CT-2006-035863 ``UniverseNet'' 
and SFB-Transregio 33 "The Dark Universe" by Deutsche
Forschungsgemeinschaft (DFG).
\end{theacknowledgments}


\begin{thebibliography}{48}
\expandafter\ifx\csname natexlab\endcsname\relax\def\natexlab#1{#1}\fi
\providecommand{\enquote}[1]{``#1''}
\expandafter\ifx\csname url\endcsname\relax
  \def\url#1{\texttt{#1}}\fi
\expandafter\ifx\csname urlprefix\endcsname\relax\def\urlprefix{URL }\fi
\providecommand{\eprint}[2][]{\url{#2}}

\bibitem[Georgi and Glashow(1974)]{Georgi:1974sy}
H.~Georgi, and S.~L. Glashow, \emph{Phys. Rev. Lett.} \textbf{32}, 438--441
  (1974).

\bibitem[Pati and Salam(1974)]{Pati:1974yy}
J.~C. Pati, and A.~Salam, \emph{Phys. Rev.} \textbf{D10}, 275--289 (1974).

\bibitem[Georgi(1974)]{Georgi:1975qb}
H.~Georgi  (1974), in: Particles and Fields 1974, ed. C.~E.~Carlson (AIP, NY,
  1975) p. 575.

\bibitem[Fritzsch and Minkowski(1975)]{Fritzsch:1974nn}
H.~Fritzsch, and P.~Minkowski, \emph{Ann. Phys.} \textbf{93}, 193--266 (1975).

\bibitem[Kawamura(2001)]{Kawamura:2000ir}
Y.~Kawamura, \emph{Prog. Theor. Phys.} \textbf{105}, 691--696 (2001),
  \eprint{hep-ph/0012352}.

\bibitem[Asaka et~al.(2001)]{Asaka:2001eh}
T.~Asaka, W.~Buchm{\"u}ller, and L.~Covi, \emph{Phys. Lett.} \textbf{B523},
  199--204 (2001), \eprint{hep-ph/0108021}.

\bibitem[Asaka et~al.(2003)]{Asaka:2003iy}
T.~Asaka, W.~Buchm{\"u}ller, and L.~Covi, \emph{Phys. Lett.} \textbf{B563},
  209--216 (2003), \eprint{hep-ph/0304142}.

\bibitem[Dixon et~al.(1985)]{Dixon:1985jw}
L.~J. Dixon, J.~A. Harvey, C.~Vafa, and E.~Witten, \emph{Nucl. Phys.}
  \textbf{B261}, 678--686 (1985).

\bibitem[Dixon et~al.(1986)]{Dixon:1986jc}
L.~J. Dixon, J.~A. Harvey, C.~Vafa, and E.~Witten, \emph{Nucl. Phys.}
  \textbf{B274}, 285--314 (1986).

\bibitem[Ib{\'a}{\~n}ez et~al.(1987{\natexlab{a}})]{Ibanez:1987sn}
L.~E. Ib{\'a}{\~n}ez, J.~E. Kim, H.~P. Nilles, and F.~Quevedo, \emph{Phys.
  Lett.} \textbf{B191}, 282--286 (1987{\natexlab{a}}).

\bibitem[Kobayashi et~al.(2005)]{Kobayashi:2004ya}
T.~Kobayashi, S.~Raby, and R.-J. Zhang, \emph{Nucl. Phys.} \textbf{B704}, 3--55
  (2005), \eprint{hep-ph/0409098}.

\bibitem[Buchm{\"u}ller et~al.(2006)]{Buchmuller:2005jr}
W.~Buchm{\"u}ller, K.~Hamaguchi, O.~Lebedev, and M.~Ratz, \emph{Phys. Rev.
  Lett.} \textbf{96}, 121602 (2006), \eprint{hep-ph/0511035}.

\bibitem[Kim et~al.(2007)]{Kim:2007mt}
J.~E. Kim, J.-H. Kim, and B.~Kyae, \emph{JHEP} \textbf{06}, 034 (2007),
  \eprint{hep-ph/0702278}.

\bibitem[Bailin and Love(1999)]{Bailin:1999nk}
D.~Bailin, and A.~Love, \emph{Phys. Rept.} \textbf{315}, 285--408 (1999).

\bibitem[Vaudrevange(2008)]{Vaudrevange:2008sm}
P.~K.~S. Vaudrevange  (2008), \eprint{0812.3503}.

\bibitem[Ramos-S{\'a}nchez(2009)]{RamosSanchez:2008tn}
S.~Ramos-S{\'a}nchez, \emph{Fortschr. Phys.} \textbf{57}, 907--1036 (2009),
  \eprint{0812.3560}.

\bibitem[Faraggi(1999)]{Faraggi:1997dc}
A.~E. Faraggi, \emph{Int. J. Mod. Phys.} \textbf{A14}, 1663--1702 (1999),
  \eprint{hep-th/9708112}.

\bibitem[F{\"o}rste et~al.(2004)]{Forste:2004ie}
S.~F{\"o}rste, H.~P. Nilles, P.~K.~S. Vaudrevange, and A.~Wingerter,
  \emph{Phys. Rev.} \textbf{D70}, 106008 (2004), \eprint{hep-th/0406208}.

\bibitem[Beasley et~al.(2009)]{Beasley:2008kw}
C.~Beasley, J.~J. Heckman, and C.~Vafa, \emph{JHEP} \textbf{01}, 059 (2009),
  \eprint{0806.0102}.

\bibitem[Nibbelink et~al.(2009)]{Nibbelink:2009sp}
S.~G. Nibbelink, J.~Held, F.~Ruehle, M.~Trapletti, and P.~K.~S. Vaudrevange,
  \emph{JHEP} \textbf{03}, 005 (2009), \eprint{0901.3059}.

\bibitem[Ib{\'a}{\~n}ez et~al.(1987{\natexlab{b}})]{Ibanez:1986tp}
L.~E. Ib{\'a}{\~n}ez, H.~P. Nilles, and F.~Quevedo, \emph{Phys. Lett.}
  \textbf{B187}, 25--32 (1987{\natexlab{b}}).

\bibitem[Susskind(2003)]{Susskind:2003kw}
L.~Susskind  (2003), \eprint{hep-th/0302219}.

\bibitem[Schellekens(2006)]{Schellekens:2006xz}
A.~N. Schellekens  (2006), \eprint{physics/0604134}.

\bibitem[L{\"u}st(2007)]{Lust:2007kw}
D.~L{\"u}st  (2007), \eprint{0707.2305}.

\bibitem[Lebedev et~al.(2007{\natexlab{a}})]{Lebedev:2006kn}
O.~Lebedev, et~al., \emph{Phys. Lett.} \textbf{B645}, 88--94
  (2007{\natexlab{a}}), \eprint{hep-th/0611095}.

\bibitem[Lebedev et~al.(2008{\natexlab{a}})]{Lebedev:2007hv}
O.~Lebedev, et~al., \emph{Phys. Rev.} \textbf{D77}, 046013
  (2008{\natexlab{a}}), \eprint{0708.2691}.

\bibitem[Lebedev et~al.(2008{\natexlab{b}})]{Lebedev:2008un}
O.~Lebedev, H.~P. Nilles, S.~Ramos-S{\'a}nchez, M.~Ratz, and P.~K.~S.
  Vaudrevange, \emph{Phys. Lett.} \textbf{B668}, 331--335 (2008{\natexlab{b}}),
  \eprint{0807.4384}.

\bibitem[Buchm{\"u}ller et~al.(2007{\natexlab{a}})]{Buchmuller:2006ik}
W.~Buchm{\"u}ller, K.~Hamaguchi, O.~Lebedev, and M.~Ratz, \emph{Nucl. Phys.}
  \textbf{B785}, 149--209 (2007{\natexlab{a}}), \eprint{hep-th/0606187}.

\bibitem[Lebedev et~al.(2007{\natexlab{b}})]{Lebedev:2006tr}
O.~Lebedev, et~al., \emph{Phys. Rev. Lett.} \textbf{98}, 181602
  (2007{\natexlab{b}}), \eprint{hep-th/0611203}.

\bibitem[Dimopoulos et~al.(1982)]{Dimopoulos:1981dw}
S.~Dimopoulos, S.~Raby, and F.~Wilczek, \emph{Phys. Lett.} \textbf{B112}, 133
  (1982).

\bibitem[Buchm{\"u}ller et~al.(2007{\natexlab{b}})]{Buchmuller:2007zd}
W.~Buchm{\"u}ller, K.~Hamaguchi, O.~Lebedev, S.~Ramos-S\'anchez, and M.~Ratz,
  \emph{Phys. Rev. Lett.} \textbf{99}, 021601 (2007{\natexlab{b}}),
  \eprint{hep-ph/0703078}.

\bibitem[Hosteins et~al.(2009)]{Hosteins:2009xk}
P.~Hosteins, R.~Kappl, M.~Ratz, and K.~Schmidt-Hoberg, \emph{JHEP} \textbf{07},
  029 (2009), \eprint{0905.3323}.

\bibitem[Kobayashi et~al.(2007)]{Kobayashi:2006wq}
T.~Kobayashi, H.~P. Nilles, F.~Pl{\"o}ger, S.~Raby, and M.~Ratz, \emph{Nucl.
  Phys.} \textbf{B768}, 135--156 (2007), \eprint{hep-ph/0611020}.

\bibitem[Araki et~al.(2008)]{Araki:2008ek}
T.~Araki, et~al., \emph{Nucl. Phys.} \textbf{B805}, 124--147 (2008),
  \eprint{0805.0207}.

\bibitem[Ko et~al.(2007)]{Ko:2007dz}
P.~Ko, T.~Kobayashi, J.-h. Park, and S.~Raby, \emph{Phys. Rev.} \textbf{D76},
  035005 (2007), \eprint{0704.2807}.

\bibitem[Petersen et~al.(2009)]{Petersen:2009ip}
B.~Petersen, M.~Ratz, and R.~Schieren  (2009), \eprint{0907.4049}.

\bibitem[Dreiner et~al.(2006)]{Dreiner:2005rd}
H.~K. Dreiner, C.~Luhn, and M.~Thormeier, \emph{Phys. Rev.} \textbf{D73},
  075007 (2006), \eprint{hep-ph/0512163}.

\bibitem[F{\"o}rste et~al.(2009)]{Forste:2009xx}
S.~F{\"o}rste, H.~P. Nilles, S.~Ramos-S{\'anchez}, and P.~K.~S. Vaudrevange,
  \emph{work in progress}  (2009).

\bibitem[Weinberg(1972)]{Weinberg:1972fn}
S.~Weinberg, \emph{Phys. Rev. Lett.} \textbf{29}, 1698--1701 (1972).

\bibitem[Kappl et~al.(2009)]{Kappl:2008ie}
R.~Kappl, et~al., \emph{Phys. Rev. Lett.} \textbf{102}, 121602 (2009),
  \eprint{0812.2120}.

\bibitem[Weinberg(1978)]{Weinberg:1977ma}
S.~Weinberg, \emph{Phys. Rev. Lett.} \textbf{40}, 223--226 (1978).

\bibitem[Wilczek(1978)]{Wilczek:1977pj}
F.~Wilczek, \emph{Phys. Rev. Lett.} \textbf{40}, 279--282 (1978).

\bibitem[Peccei and Quinn(1977)]{Peccei:1977hh}
R.~D. Peccei, and H.~R. Quinn, \emph{Phys. Rev. Lett.} \textbf{38}, 1440--1443
  (1977).

\bibitem[Kim(1979)]{Kim:1979if}
J.~E. Kim, \emph{Phys. Rev. Lett.} \textbf{43}, 103 (1979).

\bibitem[Dine et~al.(1981)]{Dine:1981rt}
M.~Dine, W.~Fischler, and M.~Srednicki, \emph{Phys. Lett.} \textbf{B104}, 199
  (1981).

\bibitem[Svr{\v{c}}ek and Witten(2006)]{Svrcek:2006yi}
P.~Svr{\v{c}}ek, and E.~Witten, \emph{JHEP} \textbf{06}, 051 (2006),
  \eprint{hep-th/0605206}.

\bibitem[Choi et~al.(2007)]{Choi:2006qj}
K.-S. Choi, I.-W. Kim, and J.~E. Kim, \emph{JHEP} \textbf{03}, 116 (2007),
  \eprint{hep-ph/0612107}.

\bibitem[Choi et~al.(2009)]{Choi:2009jt}
K.-S. Choi, H.~P. Nilles, S.~Ramos-S{\'a}nchez, and P.~K.~S. Vaudrevange,
  \emph{Phys. Lett.} \textbf{B675}, 381 (2009), \eprint{0902.3070}.

\end{thebibliography}
\end{document}